\documentclass[a4paper]{article}
\usepackage{newlfont}
\usepackage{amsfonts}
\usepackage{amsmath}
\usepackage{graphicx}
\usepackage{mathrsfs}
\usepackage{setspace}
\usepackage{cases}
\usepackage{stmaryrd}
\usepackage{txfonts}

\newcommand{\R}{\mathbb{R}}

\newcommand{\Og}{\mathcal{O}}

\begin{document}
\title{\textbf{Motility of a Model Bristle-Bot: \\
a Theoretical Analysis.} }

\author{Giancarlo Cicconofri and Antonio DeSimone\thanks{\textbf{Corresponding author:} desimone$@$sissa.it} \\ SISSA, International School of Advanced Studies
\\
 Via Bonomea 265, 34136 Trieste - Italy
\\
giancarlo.cicconofri$@$sissa.it, desimone$@$sissa.it
}
\date{October 6th, 2014}
\maketitle

\begin{abstract}
Bristle-bots are legged robots that can be easily made out of a toothbrush head and a small vibrating engine. Despite their simple appearance, the mechanism enabling them to propel themselves by exploiting friction with the substrate is far from trivial. Numerical experiments on a model bristle-bot have been able to reproduce such a mechanism revealing, in addition, the ability to switch direction of motion by varying the vibration frequency. This paper provides a detailed account of these phenomena through a fully analytical treatment of the model. The equations of motion are solved through an expansion in terms of a properly chosen small parameter. The convergence of the expansion is rigorously proven. In addition, the analysis delivers formulas for the average velocity of the robot and for the frequency at which the direction switch takes place. A quantitative description of the mechanism for the friction modulation underlying the motility of the bristle-bot is also 
 provided.
\end{abstract}

\newpage

\tableofcontents

\newpage

\section{Introduction}\label{intro}
The study of motility in biological systems and in biomimetic artificial devices has attracted considerable attention in the recent literature \cite{Chi1}. Together with swimming \cite{McNeil,AL1,PNAS,JMPS,Ho}, flying \cite{McNeil,Chi2}, walking, running and hopping \cite{SIAM}, research in this field has focused on crawling gaits, those employed by moving organisms (or devices) in continuous frictional contact with a solid substrate. Both soft and hard devices have been designed in order to crawl over a surface in the presence of a directional (asymmetric) dynamic friction coefficient, creating a mechanical ratchet \cite{N1,N2,Mah}. Similarly, snakes and snake-like robots \cite{Guo,Hi,Hu} propel themselves by exploiting the frictional anisotropy they generate on a substrate thanks to the presence of  scales on their bellies. Gastropods glide over a mucus layer by generating traveling waves of localized contraction: by sliding over the rapidly contracting part and sticking in the remaining part they produce the tractions necessary for locomotion \cite{DesNo,Lai,Lau}. Caterpillars \cite{Cas} and soft robots \cite{She} can detach partially from the substrate: they move by exerting a grip on the ground with their leading limbs, pulling forward the trailing (detached) part of their bodies.
In all these systems, a periodic internal activation can lead to sustained propulsion through a variable interaction between the body of the locomotor and the environment, alternating high friction in some parts and low friction in others during one period \cite{TA}.

Vibrating legged robots provide a different, but related example of such system. They have been proposed  as  model locomotors to study the emergence of collectively organized motion \cite{GioMah}. Nevertheless the study of their individual propulsion mechanism still offers many interesting and challenging questions. It has been suggested \cite{GioMah} that net displacements come from the modulation of friction in time due to the oscillations of the normal forces,  leading to a stick-slip motion of their feet. A bristle-bot would move forward during the stick phase, which  occurs because of the larger frictional forces caused by the robot pushing more forcefully downwards during one phase of its vertical oscillations. When such oscillations causes a decrease of the vertical pushing, then the frictional force is reduced and the robot feet slip on the ground. This results in a much smaller horizontal force in the backward direction; the periodic vertical oscillations are then accompanied by a net forward 
 displacement. DeSimone and Tatone \cite{DesTa} have proposed a simplified model to study this mechanism, in which the tangential frictional force is given by 
\begin{displaymath}
	T =  -\mu N \dot{X}
\end{displaymath} 
where $N$ is the normal reaction force exerted by the (rigid) substrate, $\dot{X}$ is the foot velocity and $\mu$ is a phenomenological proportionality constant. A striking observation in \cite{DesTa} is that the robot may be able to switch direction of motion by tuning the frequency of the engine powering the vertical oscillations. The goal of this paper is to investigate this issue and the whole propulsive mechanism of bristle-bots in detail. 

Through a full analytical treatment of the bristle-bot model, we are able to provide an approximate expression for the average velocity and an explicit formula for the inversion frequency, namely,
\begin{equation}
		\Omega_{\textrm{inv}} = \sqrt{\frac{k}{M}} \Big/ L \cos \alpha \label{invform}
\end{equation}
where $M$ is the total mass of the robot, $L$ is the length of the legs, $\alpha$ is their rest angle and $k$ is the rotational stiffness of the spring joining the legs to the robot's body (see Figure 1). As for the average velocity $\bar{v}$, we prove that the foot velocity $\dot{X}$ stabilizes after an initial transient, getting close to a periodic function given by the sum $\dot{X}  \simeq  \bar{v} + \dot{X}_{\textrm{osc}}$ where
\begin{equation}
		\bar{v} \simeq -  \frac{1}{\bar{N}}  \fint N  \dot{X}_{\textrm{osc}}  \, , \label{vbar}
\end{equation}
with $\fint$ denoting time average, and $\bar{N}$ being the average value of the normal force $N$ (which is also close to a periodic function). Formula \eqref{vbar} puts in a quantitative framework the stick-slip picture. Indeed, the average velocity $\bar{v}$ proves to be the negative of a weighted average of $\dot{X}_{\textrm{osc}}$, the feet velocity relative to $\bar{v}$, the weight being the reactive normal force $N$ transmitted by the ground during the oscillations. Therefore, in order to move, say, forward, the robot legs exploit a stronger grip due to a larger normal force when sliding backwards, and then recover when $N$ is smaller. 

The argument above explains why the average velocity of the robot may be nonzero. 
The question of determining the actual direction of motion,
i.e., the sign of $\bar{v}$, is more subtle and depends, as \eqref{vbar} indicates, on the relative phase between the oscillations $\dot{X}_{\textrm{osc}}$ of the feet and of the normal force $N$.  This is discussed in detail in Section \ref{discuss}.

The rest of the paper is organized as follows. We set up the equations of motion in Section \ref{model} and solve them formally through an asymptotic expansion in Section \ref{formal}. Therein we calculate the first three orders of such expansion, obtaining \eqref{invform} and the expression for the approximate average velocity. The convergence of our asymptotic solution, together with its regularity, periodicity and stability are analyzed in the Appendix. In Section  \ref{discuss} we derive (\ref{vbar}) and provide a quantitative description of the locomotion process.

\section{The model}\label{model}

We consider the robot legs as massless and rigid, joined to the body with a rotational spring of stiffness $k$, while we assume that their feet are in frictional contact with the substrate. The system is driven by a force $F_{\Omega}$ internal to the body coming from a mass oscillating vertically at frequency $\Omega$. For simplicity, we assume that rotations of the body are not allowed and that the legs are always in contact with the substrate. So the only degrees of freedom in our model are the horizontal coordinate of the body $u$, and the deviation $\varphi$ from the rest angle $\alpha$ that the legs form with the vertical direction. Balancing all forces we end up with the following equations of motion
{\setlength\arraycolsep{2pt}
\begin{eqnarray}
	M\ddot{h} & = & N(t) - M g + F_{\Omega}(t) \label{one} 
	\\
	\nonumber \\
	k \, \varphi & = & N(t)L\sin(\alpha + \varphi) - \mu N(t)\dot{X}L\cos(\alpha + \varphi) \label{two}
	\\	
	\nonumber \\
	M \ddot{u} & = & -\mu N(t)\dot{X} \label{three}
\end{eqnarray}}where $N$ is the normal reaction force exerted by the (rigid) substrate, $M$ is the body mass, $L$ is the length of the legs while
\begin{displaymath}
	h =  L\cos(\alpha + \varphi) \quad \textrm{and} \quad X = u + L\sin(\alpha + \varphi) \, .
\end{displaymath}

\begin{figure}
	\centering
		\includegraphics[width=0.85\textwidth]{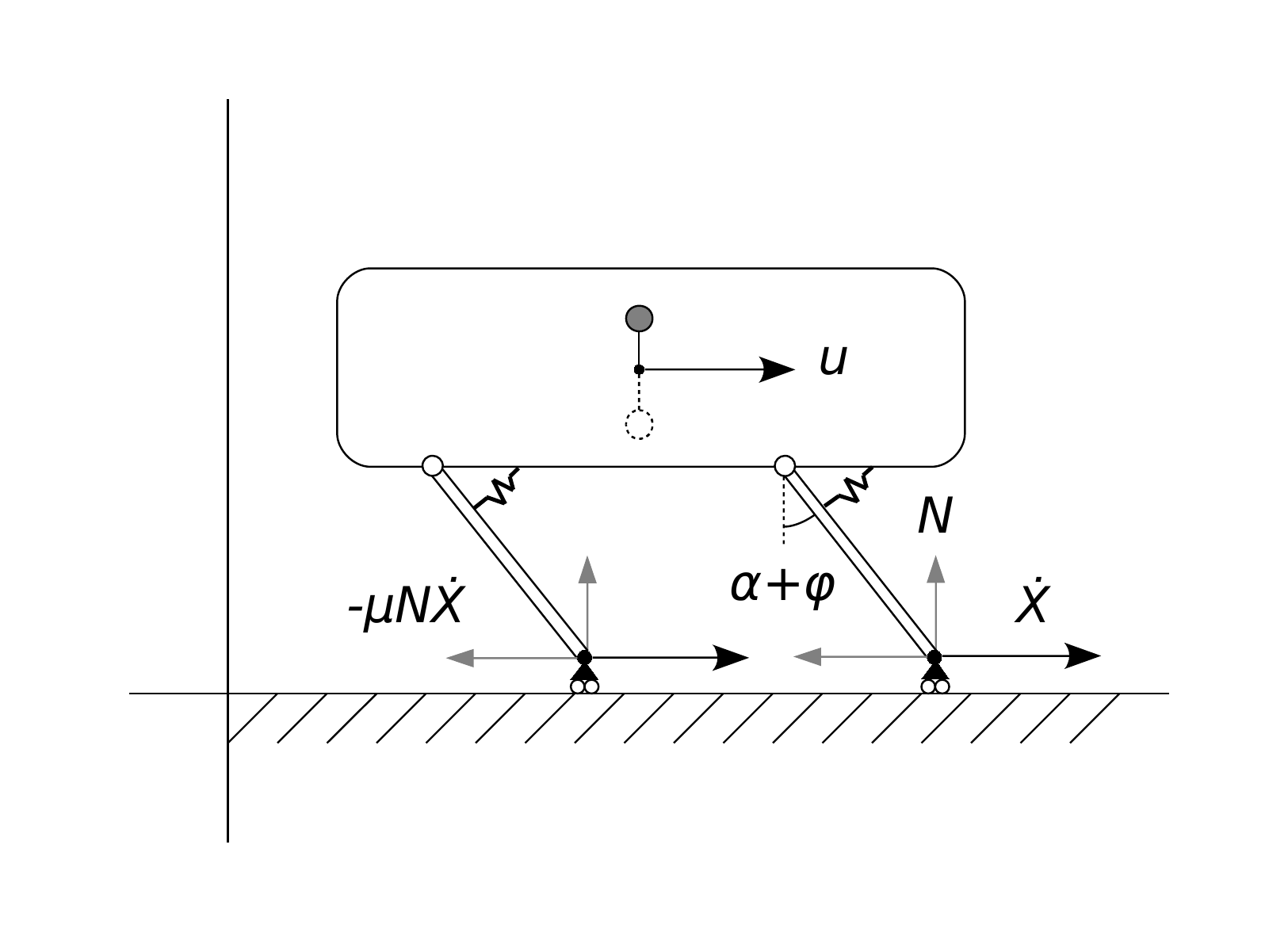}
	\caption{Schematic description of the model bristle-robot.}
 \end{figure}

We will first discuss a heuristic approach to solve the problem using an asymptotic expansion in terms of a small parameter. Instead of solving the problem directly, we will pursue the following strategy: we first give an ansatz on $N$  by choosing it in a suitable family of oscillatory functions depending on parameters. Then we find an asymptotic solution to equations (\ref{two}) and (\ref{three}) for the variables $\varphi$ and $v=\dot{u}$, that will depend on our choice of $N$. We then obtain the expression for $F_{\Omega}$ from (\ref{one}), and we find the appropriate $N$, after tuning the parameters, in order to have an approximate solution to the system in the case when the robot is driven by an oscillating internal force.

\section{Formal asymptotics}\label{formal}
Let us use the following ansatz for the normal force
\begin{displaymath}
	N(t)  =   N^{*} \! + \widetilde{N}\sin \Omega t + (N_{2}^{c}\cos2\Omega t + N_{2}^{s}\sin2\Omega t)   +  \textrm{o.h.}
\end{displaymath}
Here $N^{*}$ stands for the approximate average of the normal force, for which we take 
\begin{displaymath}
	N^{*}=Mg \, .
\end{displaymath}
This choice will be justified by the results in Section \ref{discuss}. The term $\textrm{o.h.}$ stands for ``other harmonics'' of any order, which can be neglected at first approximation. Specifically, we are considering the normalized normal force $n$, where 
\begin{displaymath}
	N(t)  =  N^{*} \! n(\Omega t) \, ,
\end{displaymath}
to be a power series expansion in the parameter $\eta$ in which the first three orders are given
\begin{equation}
	n(\tau) = 1 + \eta \sin \tau + \eta^{2}(n_{2}^{c}\cos2\tau + n_{2}^{s}\sin2\tau)  + \Og(\eta^{3}) \, . \label{nexp}
\end{equation}
The coefficients $(n_{2}^{c},n_{2}^{s})$ are the tuning parameters that will be chosen appropriately later, while we assume that $\eta$, namely the ratio between the amplitude of the first ``relevant'' harmonic and the average normal force, is a small parameter
\begin{displaymath}
\frac{\widetilde{N}}{N^{*} \! \!} = \eta \ll 1 \, . 
\end{displaymath}
\subsection{Non-dimensionalization and orders of magnitude of \\ the parameters}
We now normalize the dynamical variables, which can also be expanded into power series of $\eta$, as show below. By defining the constants
\begin{displaymath}
\sigma = \sin(\alpha) \quad \textrm{and} \quad \chi = \cos(\alpha)
\end{displaymath}
together with the angle $\epsilon$ given by
\begin{equation}
	 \epsilon = \frac{N^{*} \! L \, \sigma}{k} \label{epsilon}
\end{equation}
we determine the new dynamical variables $(\theta,w)$ through the  equalities 
\begin{equation}
	\varphi(t) = \epsilon \, \theta(\Omega t) \quad \textrm{and} \quad v(t) = \epsilon L \chi \Omega \,  w(\Omega t)\, . \label{thetaw}
\end{equation}
Applying all the definitions above we can rewrite equations (\ref{two}) and (\ref{three}) as the equivalent system
\begin{equation}
\left\{
	\begin{aligned}
\theta & = \: n(\tau)\frac{\sin(\alpha + \epsilon \theta)}{\sigma} \: - \: \xi  \, n(\tau) \! \left( w + \dot{\theta} \,\frac{\cos(\alpha +\epsilon\theta)}{\chi} \right)\frac{\cos(\alpha +\epsilon\theta)}{\chi} 
	\\
	\label{twothreenorm}
	\\
	\dot{w}& =-\lambda  \, n(\tau) \! \left( w + \dot{\theta} \,\frac{\cos(\alpha  +\epsilon\theta)}{\chi} \right) 
\end{aligned} \right.
\end{equation}
where $\tau=\Omega t$ is the non-dimensionalized time, while
\begin{equation}
		\xi  = \frac{\mu N^{*} L^{2}\chi^{2}\Omega}{k}   \quad \textrm{and} \quad   \lambda = \frac{\mu N^{*} \! }{M \Omega}  \, . \label{xilambda}
\end{equation}
Finally we normalize equation (\ref{one}), obtaining
\begin{eqnarray}
	- \left(\frac{\sigma \omega}{\chi} \right)^{2} \! \ddot{\theta} \, \frac{\sin (\alpha + \epsilon\theta)}{\sigma} - \epsilon \frac{\sigma \omega^{2} \!}{\chi} \, \dot{\theta}^{2} \frac{\cos(\alpha + \epsilon\theta)}{\chi} & = & n(\tau) - 1 + f(\tau)   \nonumber
	\\
	& & \label{onenorm}
\end{eqnarray}
where $f$ and $\omega$ are, respectively, the normalized force and frequency defined by the equations
\begin{displaymath}
		F_{\Omega}(t) = N^{*} \!f(\Omega t)  \quad \textrm{and} \quad \Omega = \sqrt{\frac{k}{M}} \frac{\omega}{L \chi} \, .
\end{displaymath}
In the next section we will formally solve (\ref{twothreenorm}), by calculating the asymptotic expansions up to the second order
\begin{equation}
\theta = \theta_{0} + \eta\theta_{1} +  \eta^{2} \theta_{2} + \Og(\eta^{3}) \quad \textrm{and} \quad   w = w_{0} + \eta w_{1} + \eta^{2}w_{2} + \Og(\eta^{3})
\label{expa}
\end{equation}
by first making the following assumptions on the parameters, needed in order to enforce the separation between $\Og(1)$ quantities and smaller ones. We take
\begin{equation}
	\omega \, , \, \xi \, , \,  \lambda \, , \, \frac{\sigma}{\chi} = \Og(1)  \quad \textrm{and} \quad \epsilon = \Og(\eta^{2}) \, \label{orders} .
\end{equation}
Such a choice of orders is always possible. Indeed we can take $\sigma,\chi = \Og(1)$, provided that we exclude the cases in which the legs are either close to perpendicular or close to parallel to the body of the robot. Then we assume $\omega = \Og(1)$, that is coherent with the fact that, as we will see, these are the order of values of $\omega$ around which the inversion of the direction of motion occurs, and that this is precisely the regime we are interested in. Finally, we can first set $\epsilon$ to be of order $\Og(\eta^{2})$, and then, exploiting the fact that $\xi$ and $\lambda$  are the only parameters depending on $\mu$, we can assume that the latter is in a range of values consistent with (\ref{orders}).

We only stress here the fact that hypothesis \eqref{orders} on the order of parameters, and specifically the one on $\epsilon$, are not strictly necessary to apply the solving technique developed in this paper, but they simplify consistently the formal developments. In particular, in the case $\epsilon=\Og(\eta^{K})$ with either $K=0$ or $K=1$  a similar analysis is still possible. However, a more complicated function for $n$ instead of \eqref{nexp} should be used, making the solution of the formal asymptotics, as well as the calculations needed for proving the rigorous results, more involved.

\subsection{Asymptotic expansion}\label{expansion}
We remark again that up to now we have just rewritten equations (\ref{two}) and (\ref{three}) in the completely equivalent system (\ref{twothreenorm}). From now on, we proceed formally to find approximating solutions to our problem. In the Appendix we will provide a rigorous proof that those solutions are indeed good approximations of the solutions of the original system, by using theorems from perturbation theory of periodic ODEs. 

Taking $\epsilon = c \eta^{2}$, with $c$ being a fixed constant, we replace the series expansions (\ref{expa}) of $\theta$ and $w$ in (\ref{twothreenorm}) and develop both sides of the equations into power series with respect to $\eta$. By matching coefficients of equal powers, we end up with a sequence of systems to be solved successively. At zero-order we have 
\begin{displaymath}
\left\{
	\begin{aligned}
	\theta_{0} & = 1 \: - \: \xi \left( w_{0} + \dot{\theta}_{0} \right)
	\\
	\\
  \dot{w}_{0} & = - \, \lambda \left( w_{0} + \dot{\theta}_{0}\right) 
	\end{aligned} \right.
\end{displaymath}
We will prove in the Appendix that this equation, and the others to come, have only one periodic solution and every other solution converge asymptotically to such periodic one. We take as $(\theta_{j},w_{j})$ with $j=0,1,2,\ldots$ the only periodic solution to the problem at each order. The zero-order periodic solution is
\begin{equation}
		\theta_{0} = 1 \quad , \quad  w_{0} = 0\: . \label{zeroord}
\end{equation}
A constant solution is coherent with the fact that, at this stage, only the non-oscillating part of $n$ is affecting the dynamics. 

Proceeding with the calculation of our expansion, the first order system is
\begin{displaymath}
\left\{
	\begin{aligned}
	\theta_{1} & =  \sin \tau -  \xi \left( w_{1} + \dot{\theta}_{1} + \sin \tau(w_{0} + \dot{\theta}_{0}) \right)
	\\
	\\
  \dot{w}_{1}  & =  -\lambda \left( w_{1} + \dot{\theta}_{1}  + \sin \tau(w_{0} + \dot{\theta}_{0})  \right) 
	\end{aligned} \right.
\end{displaymath}
Notice that $\sin \tau$ is the first order term in the expansion \eqref{nexp}, and that this is the first time that the oscillating part of $n$ enters in the problem. Solving these equations, imposing that the zero-order terms be the one we just found, this time we have the non-trivial periodic solution
\begin{equation}
	\theta_{1}(\tau)= \theta_{1}^{c}\cos\tau + \theta_{1}^{s}\sin\tau \quad , \quad w_{1}(\tau)= w_{1}^{c}\cos\tau + w_{1}^{s}\sin \tau \label{oneord}
\end{equation}
where
\begin{equation}\label{coeffone}
\begin{array}{ll}
\theta_{1}^{c}= \frac{ - \xi }{1+(\xi  - \lambda  )^{2}} \, , & \theta_{1}^{s}= \frac{ 1 - \lambda (\xi  - \lambda  )}{1+(\xi  - \lambda )^{2}} \, ,\\
 \\
w_{1}^{c}= \frac{ \lambda (\xi  - \lambda  )}{1+(\xi  - \lambda  )^{2}} \, ,  & w_{1}^{s}= \frac{ - \lambda }{1+(\xi  - \lambda )^{2}} \, .\\ 
\end{array}
\end{equation}
Notice that the average velocity is still zero up to the first order, and that in order to recover a non-zero average velocity we need to calculate the next order expansion in $\eta$. We have
\begin{displaymath}
\left\{
\begin{aligned}
	\theta_{2}  =  & \: \: \theta_{0} c \frac{\chi}{\sigma} \: + \: n_{2}^{s}\sin 2\tau \: + \: n_{2}^{c}\cos 2\tau  \: + \: 2 \xi  ( w_{0} + \dot{\theta}_{0})\, \theta_{0}  c \frac{\sigma}{\chi} 
	\\
	& 
	\: - \: \xi \! \left( ( w_{2} + \dot{\theta}_{2}) +  \sin\tau ( w_{1} + \dot{\theta}_{1}) + (n_{2}^{s}\sin 2\tau + n_{2}^{c}\cos 2\tau)(w_{0} + \dot{\theta}_{0}) \right) 
	\\
	\\
  \dot{w}_{2} = & \: \: \lambda  ( w_{0} + \dot{\theta}_{0} ) \, \theta_{0} \,c \frac{\sigma}{\chi} 
	\\
	&
	\: - \: \lambda  \left( ( w_{2} + \dot{\theta}_{2}) +  \sin\tau ( w_{1} + \dot{\theta}_{1}) + (n_{2}^{s}\sin 2\tau + n_{2}^{c}\cos 2\tau )(w_{0} + \dot{\theta}_{0}) \right) 
	\end{aligned}  \right.
\end{displaymath}
The only periodic solution is, in this case
\begin{equation}
	\theta_{2}(\tau)= c \frac{\chi}{\sigma} + \theta_{2}^{c}\cos 2\tau + \theta_{2}^{s}\sin 2\tau \quad , \quad w_{2}(\tau)= w^{*} \! + w_{2}^{c}\cos 2\tau + w_{2}^{s}\sin 2 \tau \label{twoord}
\end{equation}
where
\begin{equation} \label{coeff}
	\begin{array}{l}
\theta_{2}^{c}= \frac{1}{2} \left(\frac{ (1-\theta_{1}^{s}) - \theta_{1}^{c}(2\xi - \frac{\lambda}{2} )}{1+(2\xi - \frac{\lambda}{2})^{2}}\right) 

+ \frac{ (1 - \frac{\lambda}{2}(2\xi- \frac{\lambda}{2}))n_{2}^{c} - 2\xi n_{2}^{s} }{1+(2\xi - \frac{\lambda}{2})^{2}} \: ,

\\
\\
\theta_{2}^{s}= \frac{1}{2} \left(\frac{\theta_{1}^{c} + (1-\theta_{1}^{s})(2\xi - \frac{\lambda}{2} )}{1+(2\xi - \frac{\lambda}{2})^{2}}\right)   
+ \frac{ 2\xi n_{2}^{c} + (1 - \frac{\lambda}{2}(2\xi - \frac{\lambda}{2}))n_{2}^{s} }{1+(2\xi - \frac{\lambda}{2})^{2}} \: ,
 \\
 \\
w_{2}^{c}= -\frac{1}{4} \left(\frac{w_{1}^{s} + w_{1}^{c}(2\xi - \frac{\lambda}{2} )}{1+(2\xi- \frac{\lambda}{2})^{2}}\right) + \lambda\left(\frac{ (2\xi - \frac{\lambda}{2})n_{2}^{s} - n_{2}^{c} }{1+(2\xi - \frac{\lambda}{2})^{2}}\right)  , 
 \\
\\ 
w_{2}^{s}= \frac{1}{4}\left(\frac{ w_{1}^{c} - w_{1}^{s}(2\xi- \frac{\lambda}{2} )}{1+(2\xi - \frac{\lambda}{2})^{2}}\right) - \lambda\left(\frac{ (2\xi- \frac{\lambda}{2})n_{2}^{c} + n_{2}^{s} }{1+(2\xi - \frac{\lambda}{2})^{2}}\right)
\\
\end{array}
\end{equation}
and
\begin{equation}
	w^{*} = -\frac{1}{2}\left(\frac{\xi - \lambda}{1 + (\xi - \lambda)^{2}} \right) \, . \label{meanv}
\end{equation}
This last equation provides us with an explicit formula for the approximate (normalized) average velocity, and shows how its sign depends on that of the difference between the two parameters $(\xi,\lambda)$, and ultimately on the frequency. It also allow us to calculate the frequency at which the inversion of motion occurs, namely, $\omega_{\textrm{inv}}=1$ for the normalized quantity, and
\begin{displaymath}
	\Omega_{\textrm{inv}} = \sqrt{\frac{k}{M}} \Big/ L \chi \, 
\end{displaymath}
for the dimensional one. Notice that, unlike the rest of the coefficients of the second order expansion, the average velocity does not depend on the two parameters $n_{2}^{s}$ and $n_{2}^{c}$, that can be now chosen in order to solve asymptotically equation  \eqref{onenorm}, in the case when $f$ is a sinusoidal function.

\subsection{Tuning the parameters}\label{tuning}
Just by rewriting  \eqref{onenorm} we have the following expression for the normalized force
\begin{displaymath}
	f(\tau) = 1 - n(\tau) - \left(\frac{\sigma \omega}{\chi} \right)^{2} \! \ddot{\theta} \, \frac{\sin (\alpha + \epsilon\theta )}{\sigma} - \epsilon \frac{\sigma \omega^{2} \!}{\chi} \, \dot{\theta}^{2} \frac{\cos(\alpha + \epsilon\theta )}{\chi} \, .
\end{displaymath}
Substituting the expression \eqref{nexp} we assumed for $n$ and the one we calculated for $\theta$, and then formally expanding into a power series, the second member of the previous equation becomes
\begin{displaymath}
\eta \left\{ \! \! \begin{array}{c}
\textrm{sinusoidal}
\\
\textrm{terms}
\end{array} \! \! \right\} 
+
\eta^{2} \! \! \left\{ \, \left( \frac{\sigma^{2} \omega^{2} \!\!}{\chi^{2}\!} \, \theta_{2}^{c} -n_{2}^{c} \right)\cos 2\tau  + \left( \frac{\sigma^{2} \omega^{2} \!\!}{\chi^{2}\!} \, \theta_{2}^{s} -n_{2}^{s} \right)\sin 2\tau \right\} + \Og(\eta^{3}) \, .
\end{displaymath}
Now, in order to have a sinusoidal force to within $\Og(\eta^{3})$, we must require that 
\begin{displaymath}
	 \frac{\sigma^{2} \omega^{2} \!\!}{\chi^{2}\!} \, \theta_{2}^{c} - n_{2}^{c} = 0 \quad \textrm{and} \quad  \frac{\sigma^{2} \omega^{2} \!\!}{\chi^{2}\!} \, \theta_{2}^{s} - n_{2}^{s} = 0 \, .
\end{displaymath}
Since we found that
\begin{displaymath}
	\left(\! \begin{array}{c}
		\theta_{2}^{c}
		\\
		\theta_{2}^{s}
	\end{array}\!\right) = \left(\!\begin{array}{c}
		\tilde{\theta}_{2}^{c}
		\\
		\tilde{\theta}_{2}^{s}
	\end{array} \! \right) + \Theta_{2} \! \left(\! \begin{array}{c}
		n_{2}^{c}
		\\
		n_{2}^{s}
	\end{array} \! \right) \quad \textrm{with} \quad \Theta_{2} = \left( \begin{array}{cc}
		 \frac{1 - \frac{\lambda}{2}(2 \xi - \frac{\lambda}{2})}{1 + (2 \xi - \frac{\lambda}{2})^{2}} &  \frac{-2\xi }{1 + (2 \xi - \frac{\lambda}{2})^{2}} \\
		&  \\
		 \frac{ 2\xi}{1 + (2 \xi - \frac{\lambda}{2})^{2}} &  \frac{1 - \frac{\lambda}{2}(2 \xi - \frac{\lambda}{2})}{1 + (2 \xi - \frac{\lambda}{2})^{2}} \end{array} \right) 
	\end{displaymath}
and $(	\tilde{\theta}_{2}^{c}, \tilde{\theta}_{2}^{s})$ are constants, this requirement is fulfilled if the matrix
\begin{displaymath}
	\frac{\sigma^{2} \omega^{2} \!\!}{\chi^{2}\!} \Theta_{2} - \textrm{Id}
\end{displaymath}
is invertible. As it can be easily checked, this is true under the only assumption that $\xi > 0$, which is guaranteed by its definition \eqref{xilambda}.

Finally let us analyze the oscillating force that we found. We have the following asymptotic equality
\begin{displaymath}
	f(\tau) = \eta f_{1}(\tau) + \Og(\eta^{3})
\end{displaymath}
where $f_{1}$ can be calculated to be
\begin{equation}
	f_{1}(\tau) =  \frac{\sigma^{2} \omega^{2} \!\!}{\chi^{2}\!} \, \theta_{1}^{c}\cos \tau  +  \frac{\sigma^{2} \omega^{2} \!\!}{\chi^{2}\!} \, \theta_{1}^{s} \sin \tau -\sin\tau = \:
	\omega^{2} \rho_{\omega} \sin(\tau - \phi_{\omega}) \label{f1}
\end{equation}
with
\begin{displaymath}
	\rho_{\omega} =  \frac{\sigma^{2} \!}{\chi^{2}\!} \sqrt{\big(\,\theta_{1}^{c} \,\big)^{2} +  \big( \, \theta_{1}^{s} - \frac{\chi^{2}\!}{\sigma^{2} \omega^{2} \!\!} \, \, \big)^{2}} \quad \textrm{and} \quad \phi_{\omega}= \arctan \left(\frac{ \frac{\chi^{2}\!}{\sigma^{2} \omega^{2} \!\!}  -\theta_{1}^{s}}{\theta_{1}^{c}}\right) \, .
\end{displaymath}
Now, since we consider our robot as driven by a vertically oscillating mass, the expression for the normalized force must be of the type $f(\tau)=\omega^{2}r_{0}\sin(\tau)$, with $r_{0}$  being a ($\omega$-independent) constant. In order to recover such an expression for $f$ (at least up to a $\Og(\eta^{3})$ error) we must require $\eta$ to be $\omega$-dependent by imposing
\begin{equation}
	\eta_{\omega}= \frac{r_{0}}{\rho_{\omega}}\, , \label{etaom}
\end{equation}
and considering the new time variable $\tau'=\tau - \phi_{\omega}$, where $\tau'$ can be viewed as the proper (normalized) time of the internal oscillating force \eqref{f1}, while $\tau$ is the time relative to the first order harmonic of the normal force \eqref{nexp}. Notice that both these operations do not affect the analysis we proposed. Indeed, the only requirement we imposed on $\eta$ is that of being a small parameter. We can then consider it as $\omega$-dependent and having the form \eqref{etaom} if the constant $r_{0}$ is small enough, and by eventually restricting the range of values of $\omega$ in order to have $r_{0}/\rho_{\omega} \ll 1$ for all such values. Moreover, the transformation $\tau \rightarrow \tau'$ leaves the form of the equation of motion (as well as the form of each system of equations in the asymptotic expansion) invariant, therefore all the presented results still apply. In the following, we will continue to denote the small parameter as $\eta$, without explicitly considering its dependence on $\omega$, in order to avoid complications. Also, we will keep $\tau$ as the normalized time variable of the system. 

The expressions that we found for $\theta$ and $w$ provide approximate solutions to the equations \eqref{twothreenorm}-\eqref{onenorm} which are justified, at this stage,  only through a formal argument. In the Appendix we will prove that the system \eqref{twothreenorm}-\eqref{onenorm} has a unique, asymptotically stable, periodic solution $(\theta,w)$ that can be expressed by a power series in $\eta$ whose first three orders of expansion are indeed given by \eqref{zeroord}, \eqref{oneord} and \eqref{twoord}.

\section{Discussion of the physical implications}\label{discuss}
Let us turn back to the original, dimensional, equations. From the results in the Appendix it follows that the dynamical variables $(\varphi,v)$ converge asymptotically to periodic functions, provided that their initial conditions at, say, $t=0$ are close enough to the equilibrium configuration of the non-actuated system. So, for large enough values of $t$, after the initial transient, both variables can be written in a unique way as a sum of a constant (the mean value) and an ``oscillating'' periodic function with zero average
\begin{displaymath}
 \varphi \simeq \bar{\varphi} + \varphi_{\textrm{osc}} \quad   \quad v \simeq \bar{v} + v_{\textrm{osc}} \, .
\end{displaymath}
The same thing then must hold for any other function depending on them, in particular
\begin{displaymath}
	N \simeq \bar{N} + N_{\textrm{osc}} \quad \textrm{and} \quad \dot{X} \simeq \bar{\dot{X}} + \dot{X}_{\textrm{osc}} \, .
\end{displaymath}
By looking at equation (\ref{three}) we can see that $N$ can be written as the sum of the constant weight force $Mg$, the sinusoidal function $F_{\Omega}$ and the derivative of another periodic function, which therefore has zero average.  So we have that
\begin{displaymath}
	\bar{N} = N^{*} \! = Mg \, .
\end{displaymath}
The same kind of argument shows that the second term of the last member of the equation
\begin{displaymath}
	\dot{X} = v + \dot{\varphi} L \cos (\alpha +\varphi ) \simeq  \bar{v} + \left(v_{\textrm{osc}} + \dot{\varphi} L \cos (\alpha +\varphi ) \right)  
\end{displaymath}
has also zero average. Since the representation of a periodic function as a sum of its average and of an oscillating part is unique, we have that
\begin{displaymath}
\bar{\dot{X}} =   \bar{v} \quad \textrm{and} \quad \dot{X}_{\textrm{osc}} = v_{\textrm{osc}} + \dot{\varphi} L \cos (\alpha +\varphi ) \, .
\end{displaymath}
Now we can use the asymptotic representations of the various relevant quantities in equation (\ref{three}), namely
\begin{displaymath}
	M \dot{v} = - \mu N \dot{X} \, .
\end{displaymath}
Then, by integrating both members of the last equality and taking the time-averages on an interval $\left[T,T + 2\pi / \omega\right]$, for $T$ big enough, we obtain the formula for the average velocity of the robot
\begin{equation}
	\bar{v} \simeq -  \frac{1}{\bar{N}}  \fint N  \dot{X}_{\textrm{osc}}  \: .
\label{formula}
\end{equation}
This formula shows that net forward motion is due to the oscillation of $N$, which biases the product $N  \dot{X}_{\textrm{osc}}$ and leads to non-zero average speed even though $ \dot{X}_{\textrm{osc}}$ has zero average. In physical terms, the robot moves, say, more forward than backward thanks to the stronger grip available while its feet slip backward because, at these times, the robot is pushing more forcefully downwards. What `selects' the direction of motion is therefore the relative oscillation phase between normal force $N$ and the foot velocity  $\dot{X}_{\textrm{osc}}$, the latter being the combination of the velocity of the robot's center of mass and the one of the feet with respect to the body frame. From the first order system in Section \ref{formal} we have
\begin{displaymath}
w_{1}(\tau) =	\rho_{w_{1}} \! \sin (\tau - \delta_{ w_{1}} \!)\quad \textrm{and} \quad \dot{\theta}_{1} (\tau) = \rho_{\dot{\theta}_{1}} \! \sin(\tau - \delta_{\dot{\theta}_{1}}  ) 
\end{displaymath}
where all the involved quantities $\rho_{w_{1}}$, $ \rho_{\dot{\theta}_{1}} $, $ \delta_{ w_{1}} $ and $\delta_{\dot{\theta}_{1}}$ can be deduced from \eqref{oneord} and \eqref{coeffone}. All of these quantities are frequency-dependent. The functions $w_{1}$ and $\dot{\theta}_{1}$ can be viewed, respectively, as the approximate (and normalized) center of mass velocity, and the feet velocity with respect to the body frame. Their sum enters in the first approximation of the oscillating part of $\dot{X}$ according to the following equation
\begin{displaymath}
	\dot{X}_{\textrm{osc}} = \epsilon L \chi \Omega \eta \big( \, w_{1} +   \dot{\theta}_{1} + \Og(\eta) \big) \, .
\end{displaymath}
The center of mass velocity and the feet velocity with respect to the body frame have, at first order approximation, the typical behavior of a driven damped oscillator: they both vary at the same frequency of the driving force with a frequency-dependent delay and amplitude. In order to show how these delays affect the direction of motion we must recover from  (\ref{formula}) the approximate average velocity \eqref{meanv}. Let us notice first that
\begin{displaymath}
	\bar{v} \simeq -  \epsilon L \chi \Omega \left(\fint n  ( w_{\textrm{osc}} \!  + \dot{\theta} \, )  \: \: + \: \Og(\, \eta^{3}) \right)\, ,
\end{displaymath}
where $w_{\textrm{osc}}$ is the oscillating part of $w$. Now, since sines and cosines average to zero, we have
\begin{displaymath}
	\begin{array}{rcl}
		 \fint n(\tau) ( w_{\textrm{osc}}(\tau)   + \dot{\theta}(\tau)  ) & = &  \fint \eta^{2} \sin \tau \, \big( \, \rho_{w_{1}} \! \sin (\tau - \delta_{ w_{1}} \!) + \rho_{\dot{\theta}_{1}} \! \sin (\tau - \delta_{\dot{\theta}_{1}}  )  \, \big) 
		\\
		\\
		&  & + \fint \left\{\textrm{sines and cosines}\right\} + \Og(\eta^{3})
		\\
		\\
		& = & \frac{\eta^{2}}{2}\big( \, \rho_{w_{1}} \! \cos( \delta_{ w_{1}} \!) + \rho_{\dot{\theta}_{1}} \! \cos( \delta_{\dot{\theta}_{1}}  )  \, \big) + \Og(\eta^{3} \!)
	\end{array}
\end{displaymath}
Using \eqref{coeffone} and \eqref{wstar} we can express $w^{*}$ as
\begin{equation}
	w^{*} \! : = - \frac{1}{2}(w_{1}^{s} - \theta_{1}^{c}) = -  \frac{1}{2}\big( \, \rho_{w_{1}} \! \cos( \delta_{ w_{1}} \!) + \rho_{\dot{\theta}_{1}} \! \cos( \delta_{\dot{\theta}_{1}}  )  \, \big) \, , \label{wstar}
\end{equation}
therefore (\ref{formula}) becomes
\begin{equation}
	\bar{v} \simeq \epsilon L \chi \Omega \eta^{2} \big( w^{*} + \Og(\eta) \big) \, .  \label{vapp}
\end{equation}
Formulas \eqref{wstar} and \eqref{vapp} above show that the sign of $w^{*}$, and hence of $\bar{v}$, is selected by the relative magnitude of two constants that are affected by the interplay of the two frequency-dependent delays $\delta_{ w_{1}}$ and $\delta_{\dot{\theta}_{1}} $. Both signs are possible, with positive sign prevailing in the frequency range $\left[0,\Omega_{\textrm{inv}}\right)$ and negative sign emerging in the range $\left(\Omega_{\textrm{inv}},\infty\right)$.

\section{Appendix. Existence, stability and uniqueness of a periodic solution: rigorous convergence results}\label{rig}
Let us start by writing down the normalized equations of our system
\begin{equation}
\left\{\begin{array}{l}
	 - \left(\frac{\sigma \omega}{\chi} \right)^{2} \! \ddot{\theta} \, \frac{\sin (\alpha + \epsilon\theta )}{\sigma} - \epsilon \frac{\sigma \omega^{2} \!}{\chi} \, \dot{\theta}^{2} \frac{\cos(\alpha + \epsilon\theta )}{\chi}  =  n(\tau) - 1 + f(\tau)
	\\
	\\
	\theta \: = \: n(\tau)\frac{\sin(\alpha + \epsilon \theta)}{\sigma} \: - \: \xi  \, n(\tau) \! \left( w + \dot{\theta} \,\frac{\cos(\alpha +\epsilon\theta)}{\chi} \right)\frac{\cos(\alpha +\epsilon\theta)}{\chi} 
\\
\\
\dot{w}=-\lambda  \, n(\tau) \! \left( w + \dot{\theta} \,\frac{\cos(\alpha  +\epsilon\theta)}{\chi} \right)

\end{array}		\right.
\label{sys}
\end{equation}
The function $n(\tau)$ is now a \emph{derivative} quantity, that can be written in terms of $f$,  $\theta$ and its derivatives from the first equation. On the other hand the active force $f$ is now \emph{given} and we set it to be
\begin{displaymath}
	f = \eta f_{1}
\end{displaymath}
where $f_{1}$ is given by \eqref{f1}.
\\
\\
We are going to prove that, for every $\eta$ sufficiently small, this system has one and only one $2\pi$-periodic solution, which is asymptotically stable and analytic in $\eta$. This result will put the expansion of Section \ref{formal} on firm grounds. In fact, the uniqueness of the periodic solution together with the uniqueness of the power series representation for the functions involved, guarantees that we have constructed the actual solution of our problem.

By introducing the auxiliary variable $y=\dot{\theta}$ one can rewrite (\ref{sys}) obtaining the standard system of ODEs
\begin{displaymath}
\left(	\begin{array}{c}
		\dot{\theta}
		\\
		\dot{y}
		\\
		\dot{w}
	\end{array}\right) = G_{\eta} (\theta,y,w;\tau) \, ,
\end{displaymath}
where $G_{\eta}$ is an analytic function with respect to all the variables, and it is $2\pi$-periodic in $\tau$. We first study the unperturbed case
	\begin{equation}
	\left(	\begin{array}{c}
		\dot{\theta}_{0}
		\\
		\dot{y}_{0}
		\\
		\dot{w}_{0}
	\end{array}\right) = G_{0} (\theta_{0},y_{0},w_{0};\tau) \, ,
	\label{sysunp}
	\end{equation}
or, more explicitly,
\begin{displaymath}
	 \left\{	\begin{aligned}
		\dot{\theta}_{0} & = \: y_{0}
		\\
		\\
		\dot{y}_{0}  & = \frac{\chi^{2}}{\sigma^{2}\omega^{2}}\left(1- \frac{\theta_{0}}{1-\xi(w_{0} + y_{0})}\right)
		\\
		\\
		\dot{w}_{0}  & = - \lambda \left(\frac{\theta_{0}}{1-\xi(w_{0} + y_{0})}\right)(w_{0} + y_{0})
	\end{aligned} \right .
\end{displaymath}
As we expected $G_{0}$ is independent of $\tau$ because we ruled out the oscillating force. It can be immediately checked that
\begin{displaymath}
	q_{0} : = \left(	\begin{array}{c}
		1
		\\
		0
		\\
	  0
	\end{array}\right)
\end{displaymath}
is a solution, which is coherent with the results in Section \ref{expansion}. We calculate now the Jacobian matrix $DG_{0}$ at the point $q_{0}$. This will give us information about the stability of the autonomous system and will be crucial in the proof related to existence. We have
\begin{displaymath}
	 DG_{0} (q_{0}) = \left(\begin{array}{ccc}
		0                        &           1          &              0                              \\ 
		\frac{- \chi^{2}}{\sigma^{2}\omega^{2}}     &   \frac{-\chi^{2}\xi }{\sigma^{2}\omega^{2}}     &  \frac{-\chi^{2}\xi }{\sigma^{2}\omega^{2}}        \\
		  0                      &           -\lambda   &              -\lambda
	\end{array}\right) \, .
\end{displaymath}
Therefore, the characteristic polynomial is
\begin{displaymath}
	-\det(DG_{0} (q_{0}) - x\textrm{Id}) = x^{3} + \frac{\chi^{2} (\lambda + \xi)}{\sigma^{2}\omega^{2}}\, x^{2} + x + \frac{\chi^{2}\lambda }{\sigma^{2}\omega^{2}}\, .
\end{displaymath}
We recall that, for a cubic polynomial $p(x)=p_{3}x^{3} + p_{2}x^{2}+p_{1}x + p_{0}$, in order to have all three complex roots with negative real part, it is necessary and sufficient that all the coefficients $p_{j}$ are positive and that $p_{1}p_{2}-p_{0}p_{3}>0$. Since $\lambda,\xi>0$ this holds for the characteristic polynomial of $DG_{0}(q_{0})$. As a first consequence, this proves that $q_{0}$ is a (locally) asymptotically stable solution of the unperturbed system. Nonetheless this is also a sufficient condition (see \cite{MF}, theorems $6$.$1$.$1$, $6$.$1$.$2$ and $6$.$1$.$3$) to guarantee the existence, uniqueness, periodicity and asymptotic stability of the solution of the general ($\eta$-dependent) system (\ref{sys}). We give here a sketch of the proof for the reader's convenience.
\\
\\

We have to consider the solution 
	\begin{displaymath}
\left(	\begin{array}{c}
		\theta(\tau)
		\\
		y(\tau)
		\\
		w(\tau)
	\end{array}\right) = s(q,\eta,\tau) 
\end{displaymath}
to the $\eta$-dependent problem with initial data
\begin{displaymath}
\left(	\begin{array}{c}
		\theta(0)
		\\
		y(0)
		\\
		w(0)
	\end{array}\right) = q \, .
\end{displaymath}
The general theory of ODEs guarantees that such a solution exists locally for small enough values of $\eta$ and initial data $q$ close enough to $q_{0}$ and that, for such values, it is analytic. In addition,  we also know that $s(q,\eta,\cdot)$ converges to the solution of the unperturbed system as its maximal interval of definition approaches the whole real line (since the solution to (\ref{sysunp}) with initial value close to the equilibrium ones is defined on $\R$). There are no restriction then to suppose that $s(q,\eta,\cdot)$ is defined on, say, the interval $\left[0,2\pi\right]$, for every small enough values of $\eta$. Now, one can easily check that $s(q,\eta,\cdot)$ is $2 \pi$-periodic (and therefore defined on $\R$) if and only if
\begin{displaymath}
	s(q,\eta,2\pi) - q = \left(\begin{array}{c}
		0 \\ 0 \\ 0
	\end{array}\right) \, .
\end{displaymath}
We already know that
\begin{displaymath}
		s(q_{0},0,2\pi) = q_{0} 
\end{displaymath}
since the solution of the unperturbed system is constant for the initial data $q_{0}$. To prove that there exists one and only one function
\begin{displaymath}
	\eta \mapsto q_{\eta}
\end{displaymath}
defined around $\eta=0$ and such that
\begin{displaymath}
		s(q_{\eta},\eta,2\pi) - q_{\eta} = \left(\begin{array}{c}
		0 \\ 0 \\ 0
	\end{array}\right)
\end{displaymath}
one needs to apply the implicit function theorem. We have to verify that
\begin{equation}
	\det\left(D_{q}s (q_{0},0,2\pi) - \textrm{Id}\right) \neq 0 \, .
\label{det}
\end{equation}
From its definition we know that $s(q,0,\cdot)$ is the solution to the problem
\begin{displaymath}
 \left\{
	\begin{aligned}
	\dot{s}(q,0,\tau) & =  G_{0}(s(q,0,\tau))
	\\
	s(q,0,0) & =q
	\end{aligned} \right.
\end{displaymath}
We can therefore differentiate both members of the previous equations and obtain that
\begin{displaymath}
 \left\{
	\begin{aligned}
	\frac{d}{d\tau} D_{q}s(q_{0},0,\tau) & =  DG_{0}(q_{0})D_{q}s(q_{0},0,\tau)
	\\
	D_{q}s(q_{0},0,0) & = \textrm{Id}
	\end{aligned} \right.
\end{displaymath}
From this we have
\begin{displaymath}
	D_{q}s(q_{0},0,2\pi)= e^{2\pi DG_{0}(q_{0})} \, .
\end{displaymath}
But then relation (\ref{det}) is verified since all of the eigenvalues of $DG_{0}(q_{0})$ have negative real part. Thanks again to the implicit function theorem we can conclude that the only periodic solution
\begin{displaymath}
	(\eta,\tau) \mapsto s(q_{\eta},\eta,\tau)
\end{displaymath}
to problem (\ref{sys}) is analytic in $\eta$ being the composition of analytic functions. 

The asymptotic stability of the general solution for small enough values of $\eta$, which is inherited by the asymptotic stability of the unperturbed one, follows now by applying classical theorems, see \cite{MF} (theorem $6$.$1$.$3$).
\\
\\
\textbf{Acknowledgments}. This work is part of the research project founded by the European Research Council through the Advanced Grant $340685$-MicroMotility.


\begin{thebibliography}{30}


\bibitem{McNeil}  Alexander R M N (2003). Principles of Animal Locomotion. \emph{Princeton University Press}.

\bibitem{AL1} Alouges F, DeSimone A, Giraldi L,  Zoppello M (2013). Self-propulsion of slender micro-swimmers by curvature control: N-link swimmers. \emph{International Journal of Non-Linear Mechanics}  \textbf{56} :  132-141.

\bibitem{PNAS} Arroyo M, Heltai L, Millán D,  DeSimone A (2012). Reverse engineering the euglenoid movement.\emph{ Proceedings of the National Academy of Sciences} \textbf{109.44} : 17874-17879.

\bibitem{JMPS} Arroyo M, DeSimone A (2014). Shape control of active surfaces inspired by the movement of euglenids.\emph{ Journal of the Mechanics and Physics of Solids} \textbf{62} : 99-112.

\bibitem{Cas} Casey T M (1991). Energetics of caterpillar locomotion: biomechanical constraints of a hydraulic skeleton. \emph{Science} \textbf{252.5002} : 112-114.

\bibitem{Chi1} Childress S, Hosoi A, Schultz W W,  Wang Z J (2012). Natural Locomotion in Fluids and on Surfaces: Swimming, Flying, and Sliding. The IMA Volumes in Mathematics and its Applications, no. 155. \emph{New York, NY: Springer}.

\bibitem{Chi2} Childress S (1981). Mechanics of swimming and flying. \emph{Cambridge, UK: Cambridge University Press}.

\bibitem{DesNo} DeSimone A, Guarnieri  F, Noselli G, Tatone A (2013). Crawlers in viscous environments: Linear vs nonlinear rheology. \emph{International Journal of Non-Linear Mechanics}  \textbf{56} : 142-147
\bibitem{DesTa} DeSimone A, Tatone A (2012). Crawling motility through the analysis of model locomotors: two
case studies. \emph{The European Physical Journal E} \textbf{35}: 85.
\bibitem{MF} Farkas M (1994). Periodic motions.\emph{ New York, NY: Springer-Verlag}.

\bibitem{N1} Gidoni P, Noselli G, DeSimone A (2014). Crawling on directional surfaces. \emph{International Journal of Non-Linear Mechanics} \textbf{61} : 65-73.

\bibitem{GioMah} Giomi L, Hawley-Weld N, Mahadevan L (2013). Swarming, swirling and stasis in sequestered bristle-bots. \emph{Proceedings of The Royal Society A} \textbf{469}: 20120637.
\bibitem{Guo}  Guo  Z V, Mahadevan L (2008). Limbless undulatory propulsion on land. \emph{Proceedings of the National Academy of Sciences} \textbf{105.9} : 3179-3184. 

\bibitem{Hi} Hirose S (1993). Biologically inspired robots: snake-like locomotors and manipulators. \emph{Oxford University Press}.

\bibitem{SIAM} Holmes P, Full R J, Koditschek D,  Guckenheimer J (2006). The dynamics of legged locomotion: Models, analyses, and challenges. \emph{SIAM Review} \textbf{48.2} : 207-304.

\bibitem{Ho} Hosoi A E (2011). Locomotion at  low Reynolds numbers. In: Ben Amar M  et al., eds. New Trends in the Physics and Mechanics of Biological Systems: Lecture Notes of the Les Houches Summer School: Volume 92, July 2009.  \emph{Oxford University Press}.
\bibitem{Hu} Hu  D  L, Nirody  J, Scott T,  Shelley  M  J (2009). The mechanics of slithering locomotion. \emph{Proceedings of the National Academy of Sciences} \textbf{106.25} : 10081-10085.
\bibitem{Lai} Lai J  H, del Alamo J  C, Rodríguez-Rodríguez  J,  Lasheras  J  C (2010). The mechanics of the adhesive locomotion of terrestrial gastropods. \emph{The Journal of Experimental Biology} \textbf{213.22} : 3920-3933.
\bibitem{Lau} Lauga E, Hosoi A E (2006). Tuning gastropod locomotion: Modeling the influence of mucus rheology on the cost of crawling. \emph{Physics of Fluids} \textbf{18} : 113102.

\bibitem{N2} Noselli G,  DeSimone A (2014). A robotic crawler exploiting directional frictional interactions: experiments, numerics and derivation of a reduced model. \emph{Proceedings of the Royal Society A} \textbf{ 470.2171} : 20140333.

\bibitem{Mah} Mahadevan L, Daniel S, Chaudhury M K  (2004). Biomimetic ratcheting motion of a soft, slender, sessile gel. \emph{Proceedings of the National Academy of Sciences} \textbf{101.1} : 23-26.

\bibitem{She} Shepherd R F, Ilievski F, Choi W, Morin S A, Stokes A A, Mazzeo A D, Chen X, Wang M, Whitesides G M (2011). Multigait soft robot. \emph{Proceedings of the National Academy of Sciences} \textbf{108.51} : 20400-20403.

\bibitem{TA} Tanaka Y, Ito K, Nakagaki T,  Kobayashi R (2012). Mechanics of peristaltic locomotion and role of anchoring. \emph{Journal of The Royal Society Interface} \textbf{9.67} : 222-233.


\end{thebibliography}
\end{document}